\documentclass{article}
\usepackage[utf8]{inputenc} 
\usepackage[T1]{fontenc}    
\usepackage{hyperref}       
\hypersetup{colorlinks,allcolors=black}
\usepackage{url}            
\usepackage{booktabs}       
\usepackage{amsfonts}       
\usepackage{nicefrac}       
\usepackage{microtype}      
\usepackage{lipsum}
\usepackage{graphicx}

\title{Structured light engineering using a photonic nanojet}
\author{Maryam Yousefi $^{1,\, 2}$, Daniel Ne\v{c}esal $^{1}$, Toralf Scharf $^{3}$, Markus Rossi $^{2}$ \\
        \footnotesize $^{1}$Nanophotonics and Metrology Laboratory, Ecole Polytechnique Fédérale de Lausanne (EPFL), Lausanne, 1015, Switzerland \\
        \footnotesize $^{2}$AMS AG, Moosstrasse 2, Rüschlikon, CH-8803, Switzerland \\
        \footnotesize $^{3}$SUSS MicroOptics SA, Rouges-Terres 61, CH-2068 Hauterive, Switzerland \\\\
        \footnotesize $^{*}$Corresponding author: maryam.yousefi@epfl.ch}
\begin{document}
\maketitle
\begin{abstract}
In this letter, we present the photonic nanojet as a phenomenon in a structured light generator system that is implemented to modify the source focal spot size and emission angle. The optical system comprises a microlens array that is illuminated by a focused Gaussian beam to generate a structured pattern in the far-field. By introducing a spheroid with different aspect ratios in the focus of the Gaussian beam, the source optical characteristics change and a photonic nanojet is generated which will engineer the far-field distribution. To probe the light fields we implement a high-resolution interferometry setup to extract both the phase and intensity at different planes. We both numerically and experimentally demonstrate that the pattern distribution in the far-field can be engineered by photonic nanojet. As an example, we examine prolate, sphere, and ablate geometries. An interesting finding is that depending on the spheroid geometry, a smaller transverse FWHM of a photonic nanojet with a higher diverges angle produces an increased pattern field of view at the same physical size of the optical system.
\end{abstract}

\section{Introduction}
Photonic nanojet (PNJ) is a high-intensity strongly focused light beam that is generated on the shadow side surface of dielectric microparticles under a plane wave or a Gaussian beam  illumination \cite{heifetz_photonic_2009,devilez2009three}. The focused beam creates a propagating non-evanescent electromagnetic field with a small lateral dimension that can be smaller than the diffraction limit. Moreover, PNJ can apply to a wide range of microparticle dimensions from 2$\lambda$ to 40$\lambda$ and its optical properties are governed by the structure geometry (i.e. particle size and shape) \cite{zhang_ultralong_2018,han_controllable_2014} and refractive index \cite{mao_tunable_2015,xing_side-lobes-controlled_2018}, or by modifying the source optical properties including polarization, wavelength, etc \cite{chen_photonic_2020,devilez2009three}. Because of the flexible optical characteristics, PNJ has recently received significant attention in different fields especially in super-resolution optical imaging \cite{yang_super-resolution_2016,wang_optical_2011}, sub-wavelength direct-write nanopatterning \cite{mcleod2008subwavelength}, nanolithography \cite{kim2012fabrication,jacassi2017scanning}, nanoparticle optical trapping \cite{wang_trapping_2016,li_trapping_2016}, etc. 

Another interesting field of research is structured light pattern generation that has various applications in sensing \cite{zhang2018high}, imaging \cite{geng2011structured}, etc. For this purpose, employing micro-optical (micro-lens arrays, etc) \cite{naqavi2016high,yousefi2020light} or diffractive optical elements (binary phase grating) \cite{kim2020inverse} under a Gaussian beam illumination or a plane wave are practical methods. Considering a micro-lens array (MLA) under a focused Gaussian beam, a structured pattern is generated. In this letter, we introduce the structured pattern generation for an MLA using the PNJ phenomenon instead of a focused Gaussian beam. Our research introduces the potential of PNJ for modifying the source and engineering its size and angular distribution in a microlens based array generator. 

First, we numerically report the structured pattern generation for an MLA that is illuminated by a PNJ. We compare the PNJ for three different spheroid geometries that are illuminated by a focused Gaussian beam.  Next, in the experiment part, we introduce a high-resolution interferometry setup that is adopted to record the field intensity and phase in different planes and compare them with simulations. Here, the lateral full width at half-maximum (FWHM) of PNJ is investigated as the main optical parameter.

\section{Configuration}
\begin{figure}[htp!]
\centering
\includegraphics[width=0.75\linewidth]{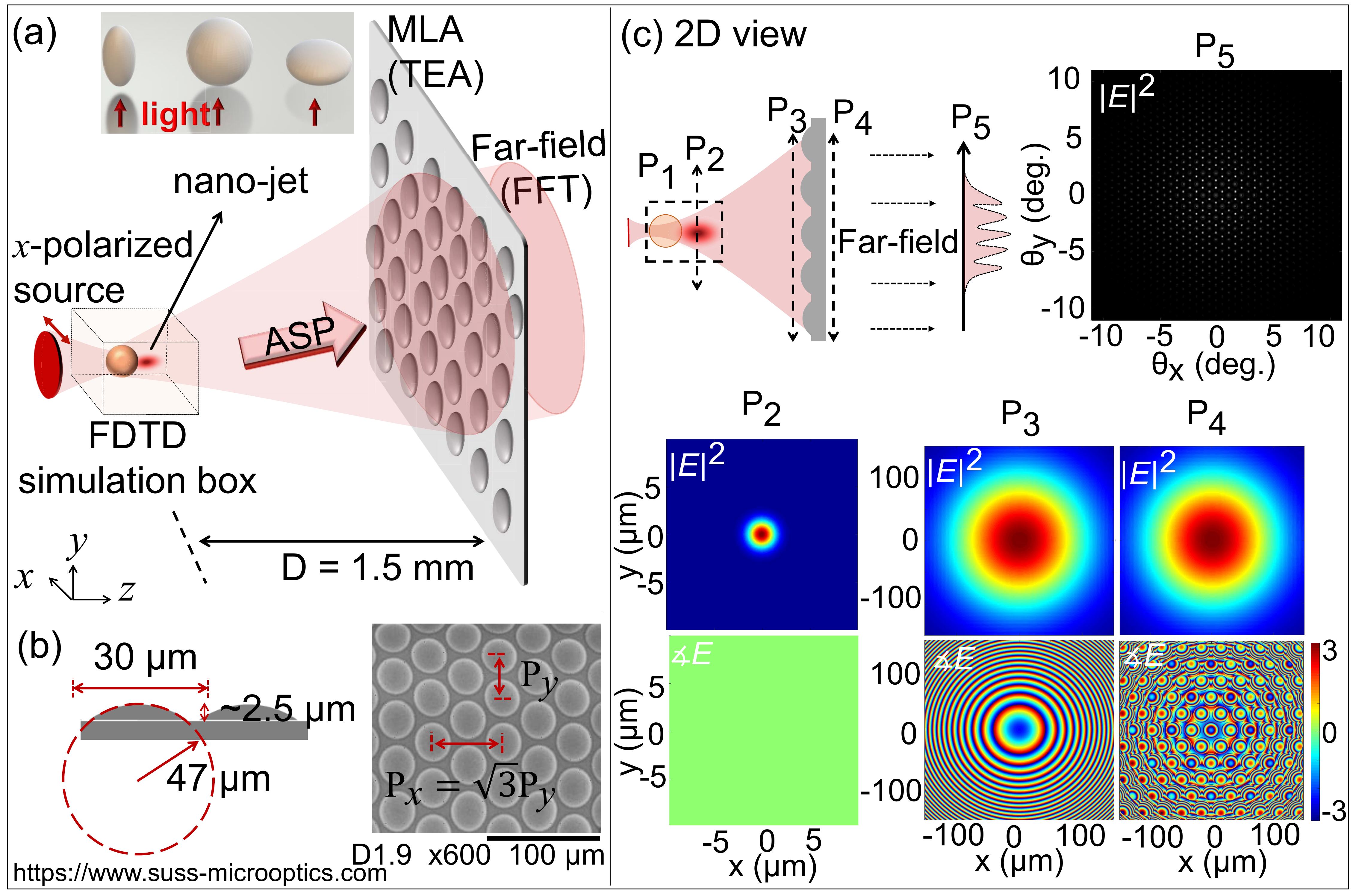}
\caption{(a) The configuration under study, (b) lens array drawing from side view and a scanning electron microscopy (SEM) image of sample from top view and (c) The MLA under the Gaussian beam illumination (no spheroid in its near-field) and the extracted fields in different planes by doing simulations.}
\label{fig:conf1}
\end{figure}

The 3$D$ schematic of our configuration is shown in Fig. \ref{fig:conf1} (a). An $x$-polarized single-mode Gaussian beam is focused on a spheroid surface, resulting in a PNJ with a high-intensity narrow peak in the shadow side. The spheroid is located at a certain distance $D$ from an MLA and the field intensity distribution is observed in the far-field. The far-field is referred to the Fraunhofer region in physical optics \cite{goodman2005introduction}. The source wavelength is $\lambda$ = 642 $nm$ in all simulations and experiments. Assuming the PNJ as a point source, for particular values of the distance $D$, a high contrast pattern is realized in the far-field based on the known self-imaging phenomenon \cite{som1990generalised}. Based on this theory, by introducing a point source, the MLA field distribution would reproduce in the far-field for certain values of the distance $D$ depending on the MLA period $P$ and source wavelength $\lambda$. Here, we choose $D$ = 1.5 $mm$ to fulfill the self-imaging condition to obtain a high contrast pattern in the far-field. More detailed discussions on the self-imaging phenomenon for a point source illumination can be found in \cite{yousefi2020near,naqavi2016high,som1990generalised} and it is beyond the scope of this paper. We examine the formation of PNJ for three spheroid geometries of the prolate, sphere, and oblate, as seen in the $3D$ schematic in inset Fig.\ref{fig:conf1} (a). The sphere diameter is 10 $\mu m$, the prolate spheroid is elongated along the $z$-axis with the dimension of 20 $\mu m$ x 10 $\mu m$,  and the oblate spheroid is flattened along the $z$-axis with the dimension of 10 $\mu m$ x 20 $\mu m$. The MLA has a hexagonal lattice with a period of 30 $\mu m$ and a lens radius of curvature of 47 $\mu m$, as seen in the SEM image in Fig. \ref{fig:conf1} (b). The MLA is made of fused silica and without aperture for each lens. The lens height is  2.5 $\mu m$ for a period of 30 $\mu m$ and for this reason, it can be considered as a thin lens.

\section{Comparison between simulations and experiments}

In the simulation part, as seen in Fig. \ref{fig:conf1}(a), the source is modeled by an $x$-polarized single-mode Gaussian beam with a beam waist of 2 $\mu m$ that propagates along the $z$-axis. To calculate the spheroid near-field, we use a $3D$ rigorous FDTD solver (Lumerical FDTD \cite{solutions_lumerical_2003}). The perfectly matched layer (PML) boundary condition is applied along the $x$, $y$, and $z$-axis with a uniform mesh size of 50 $nm$. The electromagnetic field is then extracted in the plane in which the PNJ hot spot forms. The extracted field is propagated from the PNJ to the plane just immediately before MLA for a distance of $D$ = 1.5 $mm$ by applying the angular spectrum of plane waves method (ASP) \cite{goodman2005introduction}. The effect of physical optics by passing throw the MLA is modeled by applying thin element approximation (TEA) that is valid in our case \cite{yousefi2020light} and only introduced a phase delay according to the MLA surface profile. However, no amplitude modulation is added by applying this approximation. Furthermore, as we apply a thin MLA, the pattern field of view (FOV) in the far-field does not go far beyond the paraxial approximation. For this reason, the far-field can be calculated by taking the Fourier transform of the extracted field immediately after the MLA, considering the Fraunhofer approximation \cite{goodman2005introduction}. The spheroid and the MLA refractive index are considered to be $n = 1.5$ and the whole configuration is in the air with a refractive index of one. Also, the fields are extracted in planes $P_{1}$ to $P_{5}$, as shown in the $2D$ view of our configuration in Fig. \ref{fig:conf1} (c). $P_{5}$ is the far-field plane in which we only record the pattern intensity distribution.

As reference and starting point, the field distributions at the different planes are shown for a Gaussian beam illumination (no spheroid, only Gaussian beam and MLA), in Fig. \ref{fig:conf1} (c). As seen, no phase modulation is observed in $P_{2}$ which is the Gaussian beam focal plane in this example. The beam is then propagated for a distance of $D$, resulting in both the intensity and phase modulation in the plane of $P_{3}$, immediately before the MLA. As we employ TEA, only the phase is modulated after passing through the MLA, however, the intensity distribution remains the same in planes $P_{3}$ and $P_{4}$. Finally, we observe a high contrast pattern of dots with a hexagonal distribution (because of the MLA hexagonal lattice) in the far-field with +/- $8^{\circ}$ FOV. In a similar representation, we will theoretically and experimentally study the effect of adding a spheroid with different aspect ratios in the focal plane of the Gaussian beam. In the rest of the paper, $P_{2}$ is referred to the plane in which the PNJ is formed.

\begin{figure}[htp!]
\centering
\includegraphics[width=0.75\linewidth]{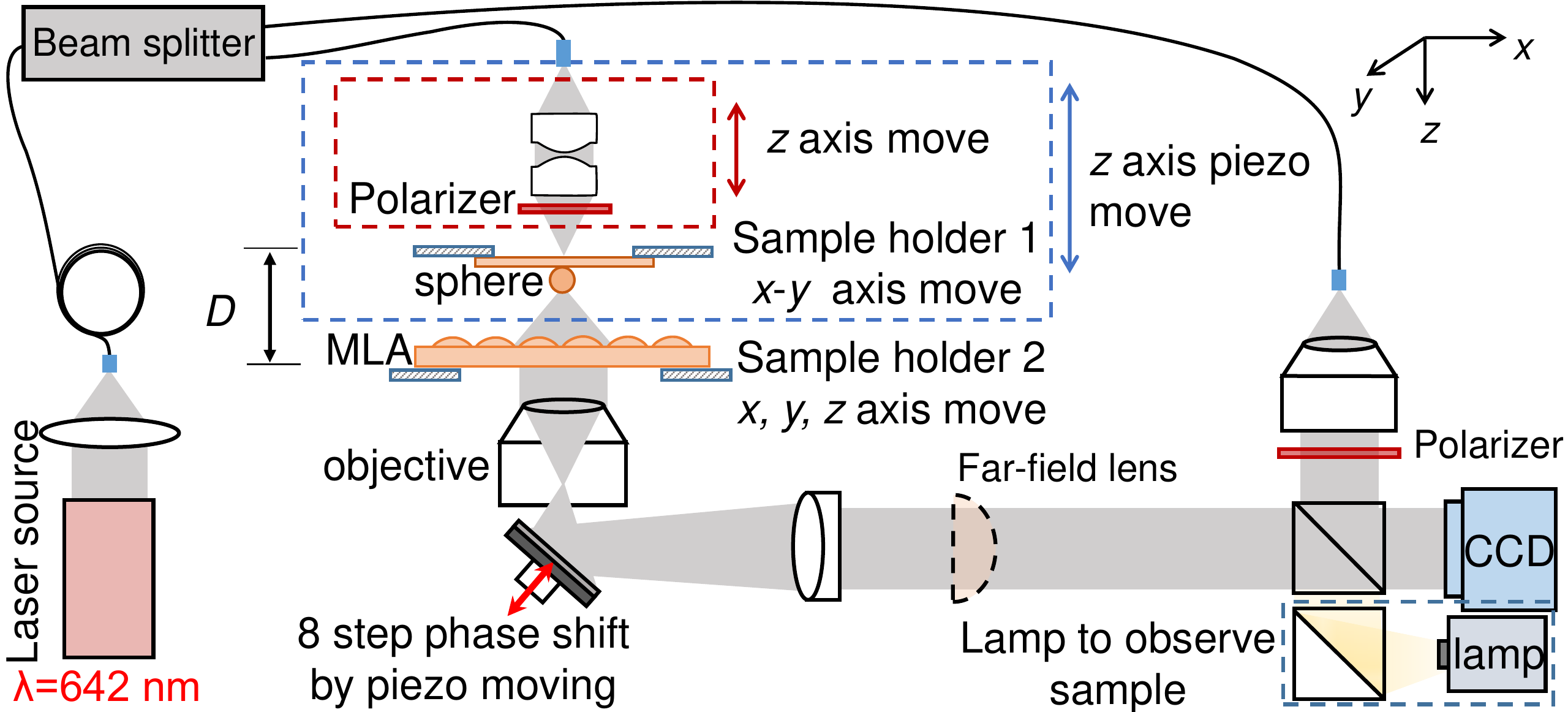}
\caption{High-resolution interferometry setup for measurement.}
\label{fig:setup}
\end{figure}

In the experimental evaluation, we aim to extract the full information of the field including intensity and phase. Although, the phase evaluation is often neglected in experiments as it adds more complexity in the optical setup.  Here, we use a high-resolution interference microscopy (HRIM) system that is a strong tool for studying the microoptical elements \cite{kim_engineering_2011}. As seen in the schematic of setup in Fig. \ref{fig:setup}, the working principle is based on the Mach-Zehnder interferometer \cite{kim_engineering_2011} that enables us to record both the intensity and phase in planes $P_1$ to $P_4$. A single-mode linearly polarized laser diode at 642 $nm$ wavelength (CrystaLaser, DL640-050-03) is utilized as the source. A beam splitter divides the intensity into the reference and object arm with the 90/10 aspect ratio. In the object arm, the beam out of the fiber is imaged using an aspheric pair lenses (C220MP-B, Thorlabs), 8 $mm$ away from the second lens surface exit. The aspheric pair lenses can freely move together along $z$ axis to precisely focus the beam on the spheroid. Sample holder 1 can freely move along $x$ and $y$ axis to precisely focus the beam at the center of the spheroid. Sample holder 2 retains the MLA and can freely move along $x$, $y$, and $z$ axis. The beam is then collected by an objective (Mitutoyo APO 50x/NA 0.42 or APO 20x/NA 0.4). The incoming beams from the reference arm and object arm interfere and are collected on the CCD camera (FLIR Point Grey, CM3-U3-50S5M-CS). By moving a piezo-electrically driven mirror, an optical path length shift is generated in the object arm. Eight interference patterns are recorded by adding an optical path length of $\lambda/4$  and the phase profile is extracted by employing the 8-step phase-shifting interferometry technique \cite{malacara2007optical}. The aspheric pair lenses and the sample holder 1 are mounted on a precision piezo stage with a $z$ scan range of 100 $\mu m$ and precision of 1 $nm$ (MadCityLabs, Nano-F100S) to extract the field in different planes.

Using the 50x objective for collecting light in the near-field of the spheroid, the lateral resolution is limited according to the objective numerical aperture (NA = 0.42) and is described by the Abbe diffraction spot size limit $\Delta x$ = $\lambda/(2 NA)$ = 764 $nm$ in the air for $\lambda$ = 642 $nm$. On the CCD camera (2048 x 2448 pixels), each pixel size corresponds to 69 $nm$ in the object plane which is smaller than the resolution limit. Employing the 20x objective to collect the MLA near-field, the field can be recorded in a large field of 422 $\mu m$ x 353 $\mu m$. The 20x objectives lateral resolution is 802 $nm$ ($\lambda$ = 642 $nm$, NA = 0.4).

Furthermore, the spheroid samples are fabricated through a direct laser writing technique by making use of the commercial Nanoscribe Photonic Professional GT which is a femtosecond laser lithography system \cite{cumpston1999two}. A few drops of IP-Dip liquid resist is dispensed on a glass wafer. Then the objective focuses on different planes to polymerize the resist with an axial and lateral resolution of 100 $nm$. After polymerization, the sample is developed and the non-polymerized resist removes. The polymerized IP-DIP refractive index is  1.52  .

\begin{figure}[ht]
\centering
\includegraphics[width=0.68\linewidth]{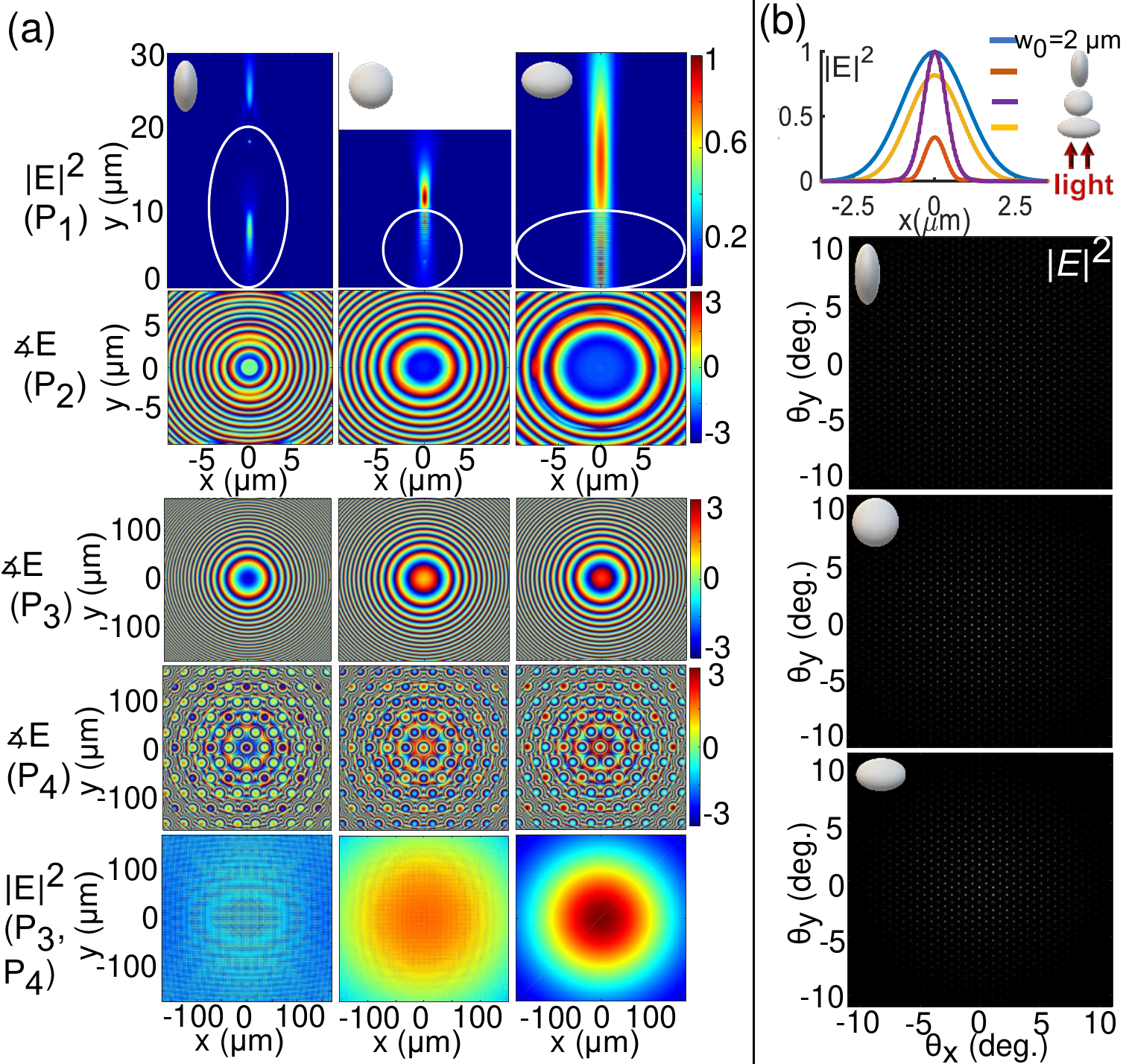}
\caption{(a) Field distribution in the planes of $P_1$ to $P_4$ for prolate, sphere and oblate, (b) The PNJ intensity cross section and the corresponding far-field distributions for simulations.}
\label{fig:sim}
\end{figure}

Figure \ref{fig:sim} illustrates the simulation results for the prolate, sphere, and oblate geometries. As seen from the spheroid nearfield in the plane of $P_1$, a hot spot is produced for all three geometries. The transverse FWHM of PNJ is calculated to be 0.76 $\mu m$ and 0.82 $\mu m$, for prolate and sphere, respectively. For oblate shape, a PNJ with a larger FWHM of 1.96 $\mu m$ in comparison with prolate and sphere is formed. One reason is that the oblate shape transverse diameter of 20 $\mu m$ is five times larger than the incoming beam diameter of 4 $\mu m$, resulting in less interference between the incoming beam and the spheroid. Besides, a PNJ with a larger transverse FWHM  and effective focal length is formed by going from a prolate to oblate geometry \cite{han_controllable_2014}. From the phase distribution in the PNJ focal plane of $P_2$, one observes that: First, the phase is modulated in the focal plane of PNJ for all the spheroid geometries while for a Gaussian beam, no phase modulation is observed at its focal plane , as seen in Fig. 1(c). Second, the smaller the transverse FWHM, the larger the phase modulations in the focus plane; i.e. that a PNJ with a higher divergence angle is produced for prolate and sphere in comparison with the oblate geometry. It also leads to different phase distributions for these configurations, in the following plane of $P_3$ immediately before the MLA. Furthermore, the phase distribution in the plane of $P_4$ immediately after the MLA is modulated according to the MLA geometry. The field intensity distribution in the plane of $P_3$ and $P_4$ immediately before and after the MLA is the same as we implement TEA for modeling the diffraction from MLA, the approximation in which no amplitude modulation is assumed. More importantly, the field intensity in the plane of $P_3$ and $P_4$ is distributed in a larger area for prolate and sphere compared to oblate shape; confirming the higher divergence angle of PNJ for prolate and sphere. 

Figure \ref{fig:sim}(b) demonstrates the transverse intensity profiles at the PNJ plane and their corresponding far-field distributions. As seen, the far-field pattern is distributed in a larger area for sphere and prolate in comparison with oblates shape; meaning that the pattern FOV is higher. It is originated from the PNJ optical characteristics. The smaller the transverse FWHM, the larger the source divergence angle, and the larger the number of lenses that are covered by the incoming beam, resulting in a higher FOV in the far-field distribution \cite{yousefi2020light}. 

\begin{figure}[ht]
\centering
\includegraphics[width=0.7\linewidth]{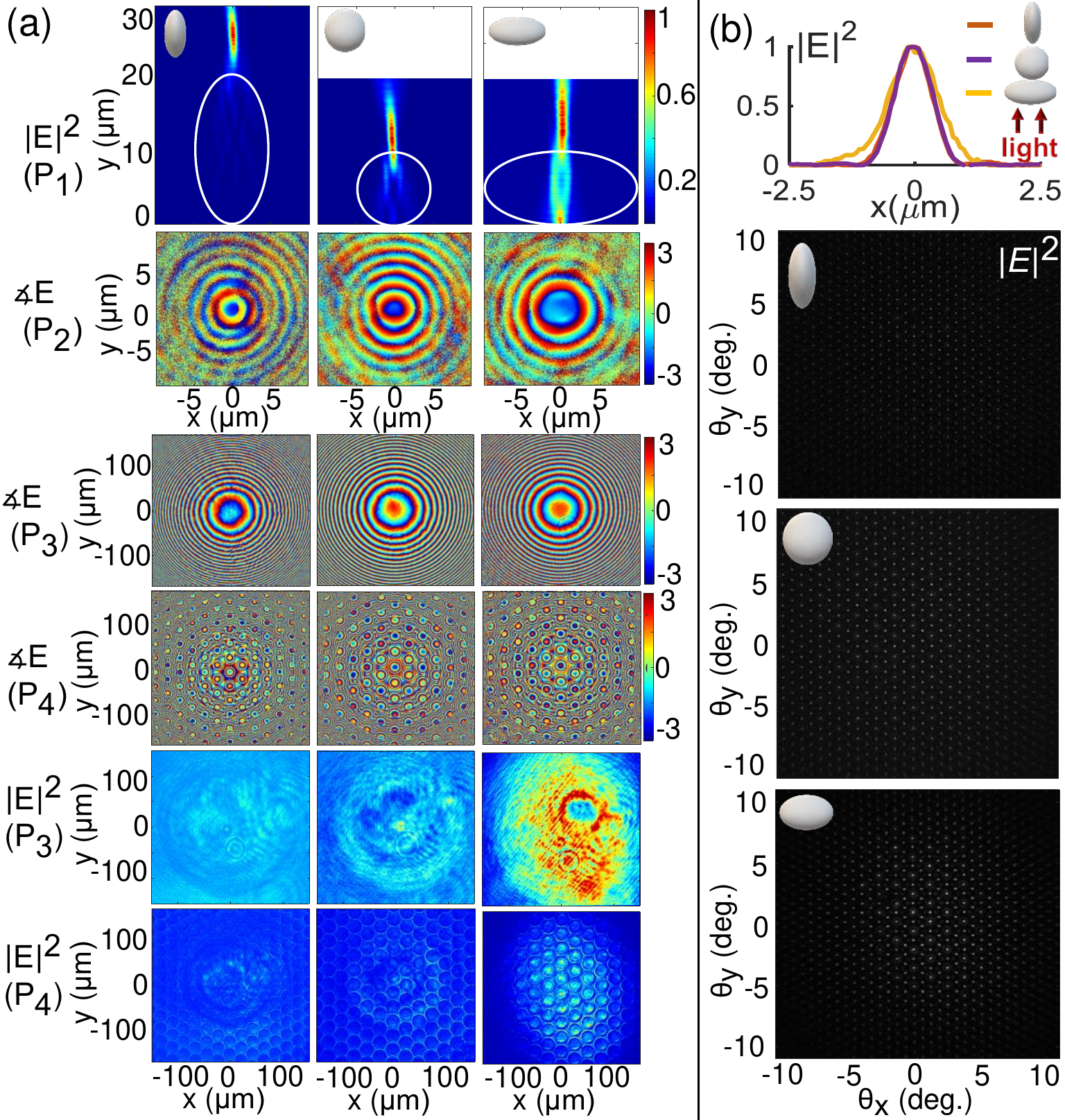}
\caption{(a) Field distribution in the planes of $P_1$ to $P_4$ for prolate, sphere and oblate
geometries, (b) The PNJ intensity cross section and their corresponding far-field distributions, by doing experiments.}
\label{fig:exp}
\end{figure}

Figure \ref{fig:exp} illustrates the experimental results for comparison with performed simulations. From the recorded near-field in the plane of $P_1$, it can be observed that a PNJ with a smaller transverse FWHM is formed for prolate and sphere in comparison with oblate shape. The transverse FWHM is measured to be 0.8 $\mu m$, 0.82 $\mu m$, and 1.06 $\mu m$ for prolate, sphere, and oblate, respectively. As seen in $P_2$, the phase distribution in the focal plane of PNJ confirmes that for smaller FWHM, more modulations are observed in the phase distribution that is analogous to a source with a higher divergence angle. However, the phase measurement in this plane is extremely challenging because the fields are mostly concentrated in the focus of PNJ and, the intensity level in the edges is very low to interfere with the incoming beam from the reference arm. For this reason, the phase distribution in outer rings is noisy. As seen, the phase distribution in $P_3$ and $P_4$, immediately before and after the MLA are recorded. In the plane of $P_4$ immediately after the MLA, the phase distribution is modulated due to the effect of diffraction from MLA. Also, the field intensity distribution is measured in $P_3$ and $P_4$ immediately before and after the MLA. As we also observed in the simulation results, the field intensity in $P_3$ immediately before the MLA is distributed in a larger area for prolate and sphere in comparison with oblate geometry. The intensity distribution in $P_4$ immediately after the MLA is weekly modulated because of the diffraction from the thin MLA. This modulation was not observed in simulation results because the TEA was employed for modeling the MLA. 

Figure \ref{fig:exp}(b) shows the transverse intensity profiles at PNJ plane and their corresponding far-field distributions. As seen, the far-field pattern is distributed in a larger area, having a higher FOV for sphere and prolate in comparison with oblates shape. For a smaller FWHM of the PNJ a higher divergence angle is introduced that results in a higher FOV in the far-field pattern for prolate and sphere geometries.

The simulation and experimental results are in good agreement although, two aspects should be considered while comparing them. First, underneath the fabricated spheroid where the structure joins the substrate is not perfectly curved because the spheroid should have flat support to attach the glass substrate \cite{bogucki2020ultra}. This effect can be more pronounced for prolate because of its high curvature in the area in which the structure is attached to the substrate. Second, the intensity is recorded by focusing the objective in different planes including the planes inside the spheroid. By focusing the objective, the interference effect inside the particle due to the material refractive index is not fully considered. 

\begin{figure}[htp!]
\centering
\includegraphics[width=0.5\linewidth]{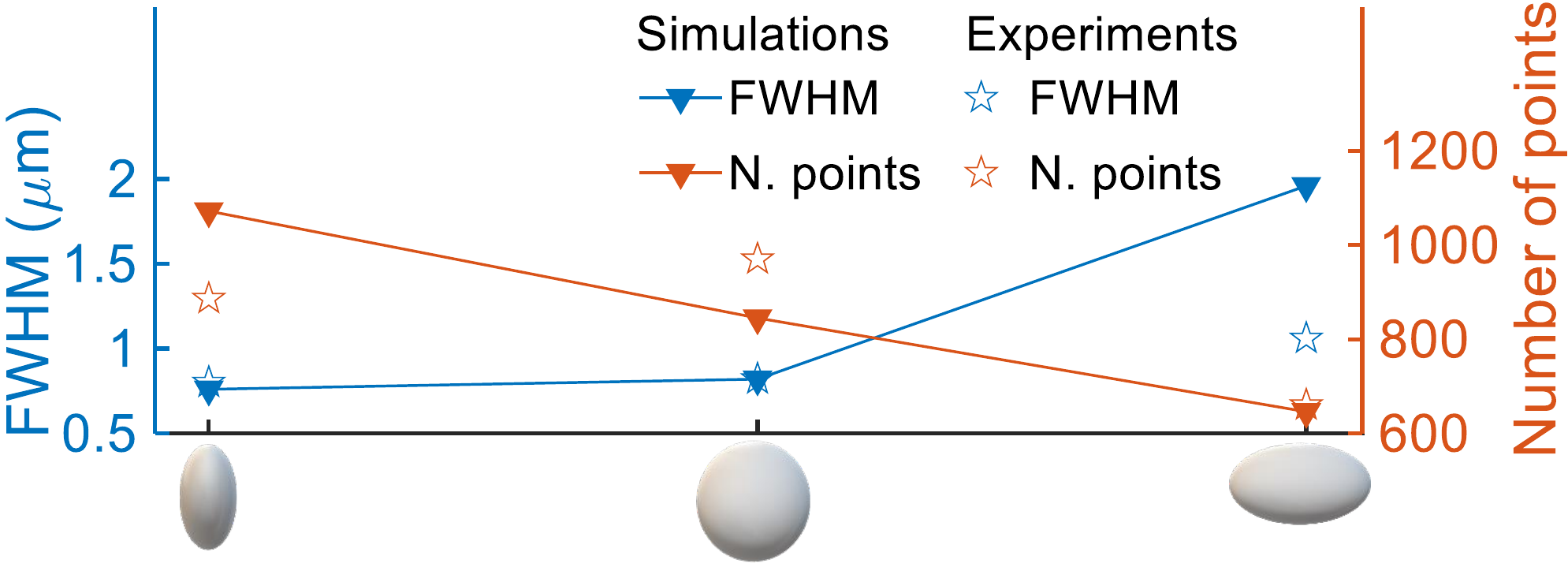}
\caption{Numerically and experimentally calculated FWHM and the number of points in the far-field pattern.}
\label{fig:simexp}
\end{figure}

For a thorough study, we do a comparison between the far-field distributions, quantitatively. We calculate the number of points in the far-field pattern considering only dots intensities higher than 13\% of the maximum intensity in each pattern. Figure \ref{fig:simexp} compares the simulated and experimentally measured transverse FWHM for each configuration and the corresponding number of points (N. of points) in the far-field pattern. As seen for both simulations and experiments, a larger number of points in the far-field pattern is realized for sphere and prolate with a smaller transverse FWHM in comparison with oblate shape.

\section{Conclusion}
In conclusion, we introduce the PNJ as a source manipulator in a structured pattern generation system. The structured light generator is a microlens array under a diverging source. By introducing a spheroid in the focal point of the source, the source near-field is modulated, producing a PNJ that affects the far-field pattern. We developed a high-resolution interferometry setup to record both the phase and intensity in different planes for a deeper comparison with simulations. The simulation and experimental results demonstrate that the number of points in the far-field pattern can be engineered by the optical characteristics of PNJ including the FWHM and divergence. We examine the prolate, sphere, and oblate geometries and demonstrate that for a smaller FWHM, a source with a higher divergence angle is realized that leads to a far-field pattern with a larger FOV.

\noindent\textbf{Funding Information}

 This research has received funding from the European Union’s Horizon 2020 research and innovation program under the Marie Skłodowska-Curie Grant Agreement No. 675745. 
 
\bigskip
\noindent\textbf{Disclosures.} The authors declare no conflicts of interest.
\bigskip


\begin{thebibliography}{1}
\bibitem{heifetz_photonic_2009}
Heifetz, Alexander, et al. "Photonic nanojets." Journal of computational and theoretical nanoscience 6.9 (2009): 1979-1992.
\bibitem{devilez2009three}
Devilez, Alexis, et al. "Three-dimensional subwavelength confinement of light with dielectric microspheres." Optics Express 17.4 (2009): 2089-2094.
 \bibitem{zhang_ultralong_2018}
Zhang, Baifu, et al. "Ultralong photonic nanojet formed by dielectric microtoroid structure." Applied optics 57.28 (2018): 8331-8337.
\bibitem{han_controllable_2014}
Han, Lu, et al. "Controllable and enhanced photonic jet generated by fiber combined with spheroid." Optics letters 39.6 (2014): 1585-1588.
\bibitem{mao_tunable_2015}
Mao, Xiurun, et al. "Tunable photonic nanojet formed by generalized Luneburg lens." Optics express 23.20 (2015): 26426-26433.
 \bibitem{xing_side-lobes-controlled_2018}
Xing, Huaming, Wenchao Zhou, and Yihui Wu. "Side-lobes-controlled photonic nanojet with a horizontal graded-index microcylinder." Optics letters 43.17 (2018): 4292-4295.
 \bibitem{chen_photonic_2020}
Chen, Ran, et al. "Photonic nanojet beam shaping by illumination polarization engineering." Optics Communications 456 (2020): 124593.
\bibitem{yang_super-resolution_2016}
Yang, Hui, et al. "Super-resolution imaging of a dielectric microsphere is governed by the waist of its photonic nanojet." Nano Letters 16.8 (2016): 4862-4870.
\bibitem{wang_optical_2011}
Wang, Zengbo, et al. "Optical virtual imaging at 50 nm lateral resolution with a white-light nanoscope." Nature communications 2.1 (2011): 1-6.
\bibitem{mcleod2008subwavelength}
Mcleod, Euan, and Craig B. Arnold. "Subwavelength direct-write nanopatterning using optically trapped microspheres." Nature nanotechnology 3.7 (2008): 413-417.
\bibitem{kim2012fabrication}
Kim, Jooyoung, et al. "Fabrication of plasmonic nanodiscs by photonic nanojet lithography." Applied Physics Express 5.2 (2012): 025201.
\bibitem{jacassi2017scanning}
Jacassi, Andrea, et al. "Scanning probe photonic nanojet lithography." ACS applied materials \& interfaces 9.37 (2017): 32386-32393.
 \bibitem{wang_trapping_2016}
Wang, Haotian, Xiang Wu, and Deyuan Shen. "Trapping and manipulating nanoparticles in photonic nanojets." Optics letters 41.7 (2016): 1652-1655.
 \bibitem{li_trapping_2016}
Li, Yuchao, et al. "Trapping and detection of nanoparticles and cells using a parallel photonic nanojet array." ACS nano 10.6 (2016): 5800-5808.
 \bibitem{zhang2018high}
Zhang, Song. "High-speed 3D shape measurement with structured light methods: A review." Optics and Lasers in Engineering 106 (2018): 119-131.
 \bibitem{geng2011structured}
Geng, Jason. "Structured-light 3D surface imaging: a tutorial." Advances in Optics and Photonics 3.2 (2011): 128-160.
 \bibitem{naqavi2016high}
Naqavi, Ali, Hans Peter Herzig, and Markus Rossi. "High-contrast self-imaging with ordered optical elements." JOSA B 33.11 (2016): 2374-2382.
 \bibitem{yousefi2020light}
Yousefi, Maryam, Toralf Scharf, and Markus Rossi. "Light pattern generation with hybrid refractive microoptics under Gaussian beam illumination." OSA Continuum 3.4 (2020): 781-797.
 \bibitem{kim2020inverse}
Kim, Dong Cheon, et al. "Inverse design and demonstration of high-performance wide-angle diffractive optical elements." Optics Express 28.15 (2020): 22321-22333.
\bibitem{goodman2005introduction}
Goodman, Joseph W. Introduction to Fourier optics. Roberts and Company Publishers, 2005.
\bibitem{som1990generalised}
Som, S. C., and A. Satpathi. "The generalised Lau effect." Journal of Modern Optics 37.7 (1990): 1215-1225.
\bibitem{yousefi2020near}
Yousefi, M., et al. "Near-field phase profile and far-field contrast modulation for micro-lens arrays at the refraction limit." SN Applied Sciences 2.12 (2020): 1-5.
\bibitem{solutions_lumerical_2003}
Solutions, F. D. T. D. "Lumerical solutions inc." Vancouver, Canada (2003).
\bibitem{kim_engineering_2011}
Kim, Myun-Sik, et al. "Engineering photonic nanojets." Optics express 19.11 (2011): 10206-10220.
\bibitem{malacara2007optical}
Malacara, Daniel, ed. Optical shop testing. Vol. 59. John Wiley \& Sons, 2007.
\bibitem{cumpston1999two}
Cumpston, Brian H., et al. "Two-photon polymerization initiators for three-dimensional optical data storage and microfabrication." Nature 398.6722 (1999): 51-54.
\bibitem{bogucki2020ultra}
Bogucki, Aleksander, et al. "Ultra-long-working-distance spectroscopy of single nanostructures with aspherical solid immersion microlenses." Light: Science \& Applications 9.1 (2020): 1-11.
\end{thebibliography}
\end{document}